\documentclass[11pt,english]{scrartcl}
\usepackage{mathptmx}
\usepackage[scaled=0.9]{helvet}
\usepackage{courier}
\usepackage[T1]{fontenc}
\usepackage[latin9]{inputenc}
\usepackage[letterpaper]{geometry}
\usepackage{amsmath}
\usepackage{amsthm}
\usepackage{setspace}
\usepackage[sort&compress, super]{natbib}
\usepackage{graphicx}
\usepackage{ulem}
\makeatletter

\newcommand{\lyxaddress}[1]{
	\par {\raggedright #1
	\vspace{1.4em}
	\noindent\par}
}

\renewcommand\[{\begin{equation}}
\renewcommand\]{\end{equation}}

\let\OLDthebibliography\thebibliography
\renewcommand\thebibliography[1]{
  \OLDthebibliography{#1}
  \setlength{\parskip}{0pt}
  \setlength{\itemsep}{0pt plus 0.3ex}
}

\makeatother

\usepackage{babel}
\begin{document}
\title{Minimal Models for RNA Simulations}
\author{D. Thirumalai$^{1}$, Naoto Hori$^{2}$, and Hung T. Nguyen$^{3}$}
\maketitle

\lyxaddress{\begin{center}
1. Department of Chemistry, The University of Texas at Austin, Austin, TX 78712, USA\\
2. School of Pharmacy, University of Nottingham, Nottingham, NG7 2RD, UK\\
3. Department of Chemistry, University at Buffalo, NY 14260, USA
\par\end{center}}

\begin{abstract}
The increasing importance of RNA as a prime player in biology can hardly be overstated. Problems in RNA, such as folding and RNA-RNA interactions that drive phase separation, require cations. Because experiments alone cannot reveal the dynamics of cation-RNA interactions, well calibrated theory and computations are needed to predict how ions control the behavior of RNA. The perspective describes the development and use of coarse-grained models at different resolutions. We focus on single- and three-interaction site interaction models, in which electrostatic interactions are treated using a combination of explicit and implicit representations. Applications to the folding of ribozymes and riboswitches are discussed, with emphasis on the role of monovalent and divalent cations. We also discuss phase separation in low complexity sequences.  Challenges in the simulation of complex problems such as ribosome assembly and RNA chaperones, requiring developments of models for RNA-protein interactions, are pointed out.
\end{abstract}

\section{Introduction}
Although RNA  sometimes referred to as the ``dark matter" of biology, the spotlight on RNA biology has never been brighter.  The award of Nobel prizes in  medicine in two consecutive years, one in 2023 for the vaccines based on nucleoside base modifications of RNA~\cite{Kariko05Immunity}, and another in 2024  for the role of microRNA in regulating gene expression~\cite{Lee93Cell,Wightman93Cell} highlight the importance of RNA. These are the latest high profile examples in RNA biology, which  took off after the demonstration that RNA could function as enzymes over forty years ago~\cite{Cech81Cell,AltmanCell83,Doudna02Nature}. In the intervening years the functional repertoire of RNA has continues to expand producing one surprise after another and with no end in sight. As is always the case in science, the discovery of new phenomena  create a treasure trove of problems that need quantitative explanations, which in the studies of RNA folding and RNA-RNA interactions require both biophysical experiments~\cite{Woodson05COSB,Woodson10ARB,Herschlag18CSHL,jain_rna_2017} and computational approaches~\cite{Hyeon05PNAS, Hayes15PRL,Hayes12JACS,Boniecki16NAR,Xia10JPCB,Nguyen22NatChem,Aierken24JCTC,Pasquali10JPCB,Sulc14JCP}.

Here,  we focus exclusively on the utility of coarse-grained (CG) models in the folding of RNA and phase separation in low complexity sequences. Although not a substitute for atomically detailed simulations in water with explicit ions~\cite{Cheatham00ARPC}, there is also a need for CG models in understanding the folding of ribozymes (RNA enzymes)   as well as in the simulations of multiple chains that span a range of time and length scales. hver the last two decades several studies have shown that CG models, beginning with the initial reports on the mechanical force-induced unfolding of a RNA hairpin~\cite{Hyeon05PNAS,Hyeon08JACS}, have been efficacious in making predictions that compare favorably with experiments. The CG models, at different resolutions, have been particularly important in elucidating the role of divalent cations in controlling RNA folding~\cite{Denesyuk15NatChem,Hori23NAR,Fuks22Frontiers,Sulc14JCP}.
It is also possible to convert a coarse-grained representation into atomic coordinates for further analysis~\cite{Perry23JMB}.

For the purposes of this article, let us classify CG models into a few categories depending on the resolution of the RNA and the treatment of ions (Figure 1).  (i) In the Single Interaction Site (SIS) model, each nucleotide is represented by a single bead~\cite{Hyeon06Structure,Regy20NAR,Guerra24JCP}. Electrostatic interactions associated with phosphate groups and the cations are modeled implicitly using the Debye-H\"{u}ckel (DH) interactions~\cite{Hyeon06Structure,Regy20NAR,Maity23PNAS}. (ii) In the Three Interaction Site (TIS) model~\cite{Hyeon05PNAS}, each nucleotide is represented using three sites corresponding to phosphate, sugar, and base. Interactions between charged groups (phosphate and metal ions) are treated explicitly~\cite{Denesyuk15NatChem} or implicitly~\cite{Denesyuk13JPCB}, depending on the applications.
(iii) Because the concentration of monovalent ions is higher than  divalent ions under both physiological and \textit{in vitro} experimental conditions, we developed a mixed description of the electrostatic interactions and combined it with the TIS model for RNA.  In the resulting TIS-CAT~\cite{Nguyen19PNAS} model, monovalent ions are treated using the DH potential whereas divalent cations (Mg\textsuperscript{2+} or Ca\textsuperscript{2+}) are explicitly treated for the accurate description of RNA folding.
(iv) There are models that use higher resolution (six or seven beads per nucleotide) in which stacking and base pairing  interactions are treated in an \textit{ad hoc} manner \cite{Pasquali10JPCB,Uusitalo17BJ,Li24bioRixv}.
In this review, our main focus is recent development of the first three categories. In addition, models that have been developed for the purpose of structure predictions\cite{Li21FrontiersMolBiosci}, while important, are beyond the scope of this review. We refer the reader to a more comprehensive review of various RNA simulations and earlier development of CG models~\cite{sponer_rna_2018}.

\begin{figure}
    \centering
    \includegraphics[width=0.75\linewidth]{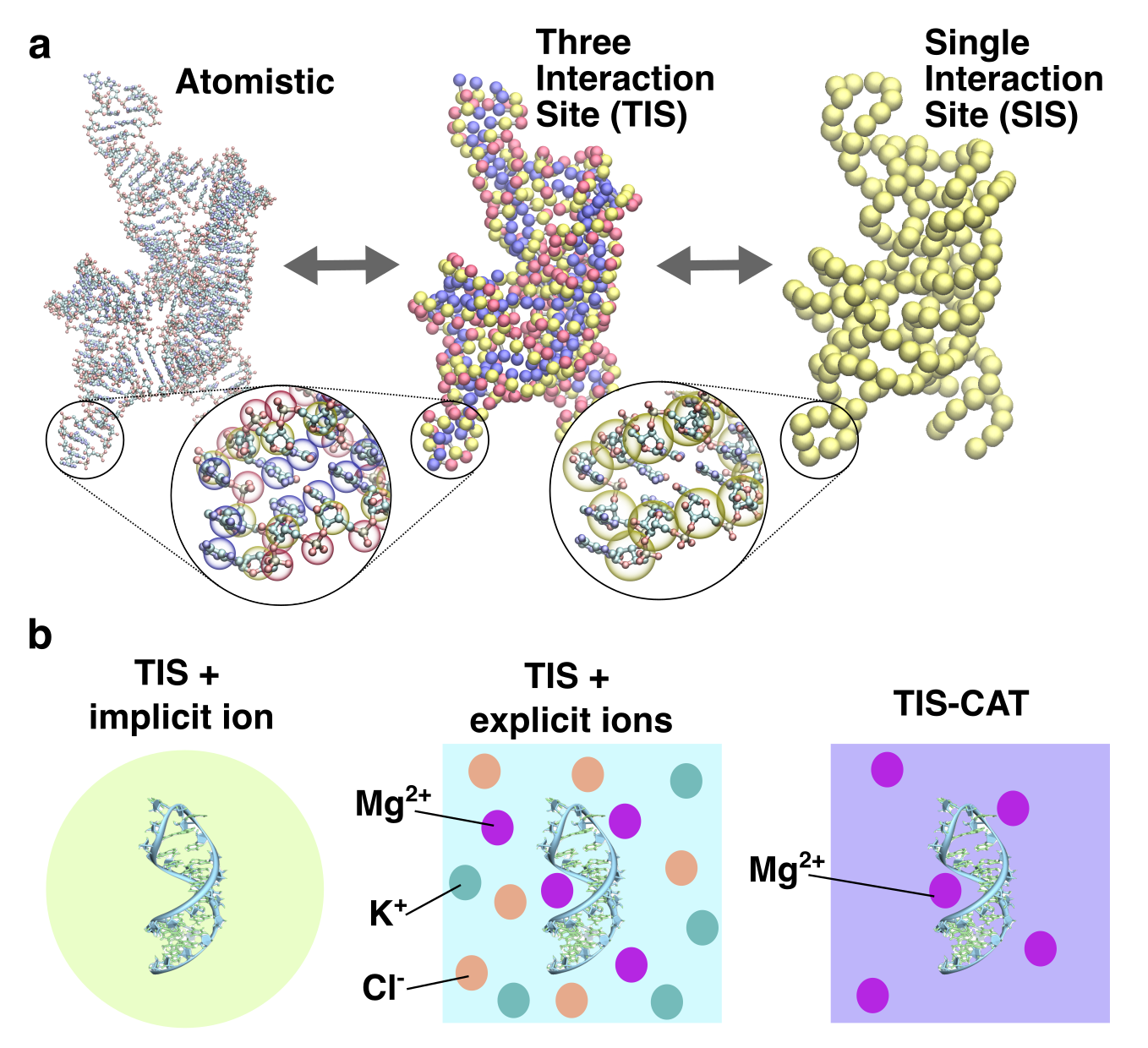}
    \caption{Coarse-grained models for RNA. (a) Hierarchy of models at different resolution. In the TIS model, each nucleotide consists of three sites representing phosphate (red), sugar (yellow), and base (blue), whereas in the SIS model, a single bead represents an entire nucleotide. (b) Treatment of ions in the TIS model framework. \textit{Left:} When the solvent contains only monovalent ions, implicit ion treatment based on Debye-H{\"u}ckel theory is appropriate. \textit{Middle:} RNA in a solution with monovalent salts and divalent cations such as Mg\textsuperscript{2+} and Ca\textsuperscript{2+}.  \textit{Right:} In the TIS-CAT model, divalent ions are treated explicitly while monovalent ions are implicit. }
    \label{fig:1}
\end{figure}

\section{Single Interaction Site (SIS) Model}
The Single Interaction Site (SIS) model represents each nucleotide by single interaction center~\citep{Hyeon06Structure}. The SIS model for folded RNA retains chain connectivity and favorable attractive interactions between sites that stabilize the native fold. The initial formulation of the SIS model provided a theoretical framework to predict the outcomes of single molecule pulling experiments in which mechanical forces probed the folding landscapes of RNA constructs and ribozymes ~\cite{Greenleaf08Science}.  

\noindent \textbf{Response of Riboswitches to Mechanical Force:}   Riboswitches, consisting of an aptamer domain (AD) and an expression platform,  contain RNA elements in the untranslated region of mRNA that allosterically regulate translation and transcription in bacteria by binding to metabolites with exquisite specificity~\cite{Serganov13Cell,Winkler05ARM}. Often transcription or translation and riboswitch folding are coupled~\cite{Watters16NSMB,Sun18JPCB,Uhm18PNAS,lin08JACS}.  The functional selectivity is best illustrated using the purine riboswitches as examples. For example, adenine (\textit{add}) riboswitch~\cite{Serganov04ChemBiol} activates translation upon binding the metabolite purine, whereas the structurally similar \textit{pbuE} adenine riboswitch~\cite{Greenleaf08Science} activates transcription in the presence of purine. Both of these are ON riboswitches, which means that translation or transcription is activated only upon binding of the metabolite. In contrast, OFF riboswitches shut down gene expression when the metabolite binds to the aptamer domain. The SAM-III riboswitch binds S-adenosylmethionine (SAM) and represses translation upon binding SAM which sequesters the Shine-Dalgarno ribosome binding sequence. Single molecule force spectroscopy  using optical tweezers, performed either by applying constant force to the ends of the RNA or by increasing the force at a constant rate, were used to assess the characteristics  of the intermediates in riboswitches~\cite{Frieda12Science,Greenleaf08Science}.

In conventional single-molecular force spectroscopy, performed in either the  force ramp or force clamp mode, only states that are populated in the accessible force range are characterized. Experiments performed using force pulses during stretch-quench cycles could uncover the hidden states through which RNA folds. 
Starting from a fully unfolded riboswitch, at a high stretching force, refolding can be initiated by quenching the force to a value less  than the critical force needed to unfold for a time period $t_q$.
If $t_q$ is less than the folding time, RNA would only access partially structured states. By repeating the process of stretching and quenching the force for different values of $t_q$ folding can be interrupted, and the states that are difficult to capture in a single or step wise quenching of force could be resolved. The SIS model simulations with the metabolite SAM showed that a spectrum of states was populated when $t_q$ was varied. Importantly, the simulations~\cite{Lin13JACS} showed that the refolding time as a function of force is accurately predicted using calculated free energy profile and Kramers rate theory~\cite{Hanggi90RMP}. 

\noindent \textbf{Phase separation in repeat RNA sequences:}  The SIS model has been extended to probe phase separation in low complexity RNA repeat sequences~\cite{La10NatRevGenetics}.  \textit{In vitro} experiments~\cite{jain_rna_2017} showed that the trinucleotide repeat sequences such as (CAG)$_n$ and (CUG)$_n$ undergo phase separation in which a dense phase coexists with a dilute phase. Interestingly, phase separation occurs only if the number of repeats $n$ exceeds a critical value.  These findings were explained using the SIS model, with a single unknown energy scale for base-pair interactions, that was adjusted  to reproduce the known structures of a short (CAG)$_2$ duplex~\cite{Nguyen22NatChem}. Multichain simulations recapitulate the length and concentration dependence of the phase separation, in quantitative agreement with the \textit{in vitro} experiments (Figure 2 a and b). The driving force for coacervate formation is the intermolecular base-pair interactions that has both an enthalpic and entropic contribution. Unexpectedly, it was found that once RNA molecules are recruited in the droplets, they undergo large conformational change, from a hairpin-like conformation in isolation to a nearly stretched state, to form an extensive network of intermolecular interactions (Figure 2c). The formation of the soft network, in turn, constrains the RNA conformational fluctuation and mobility, which was rationalized using reptation dynamics~\cite{DeGennes71JCP}. 
The SIS model has been adopted to simulate a number of repeat RNA sequences using LAMMPS, which reportedly enhances computational efficiency by speeding up the simulations by a factor of three to five \cite{Aierken24JCTC}. Like the previous work, simulations of a number of low complexity RNA sequences showed that the enhanced number of ways of forming Watson-Crick pairs render stability to the dense droplets.  

Based on the SIS model, a theory was advanced for predicting the propensity of RNA to self-associate. It was discovered that the free energy gap separating the ground state and low lying excited states in repeat sequences are small relative to $k_BT$~\cite{Maity23PNAS, Maity24JPCL}. As a result, such sequences can readily unfold, thus exposing a stretch of nucleotides that can engage in inter molecular base pairing~\cite{Roden20NatRevMolCellBiol,Langdon18AnnRevMicrobiol}, resulting in phase separation. In contrast, the free energy gap in heterogeneous sequences in which the 5$^{\prime}$ and 3$^{\prime}$ ends are in proximity are sufficiently large~\cite{Maity23PNAS} that the excited state has negligible population, which prevents unwarranted RNA-RNA interactions.

The single bead representation of RNA has recently been used to investigate  phase separations in mixture of RNA and proteins.\cite{Regy20NAR, Joseph21NatComputSci,Valdes-Garcia23JCTC} In these models, interactions of RNA residues are parameterized to reproduce condensate formation with proteins, but the capability of such models to fold RNA into complex secondary and tertiary structures has not been demonstrated. Understanding the impact of RNA in influencing phase separation in proteins will also require a consistent models of RNA as well as interactions with ions.

\begin{figure}
    \centering
    \includegraphics[width=1\linewidth]{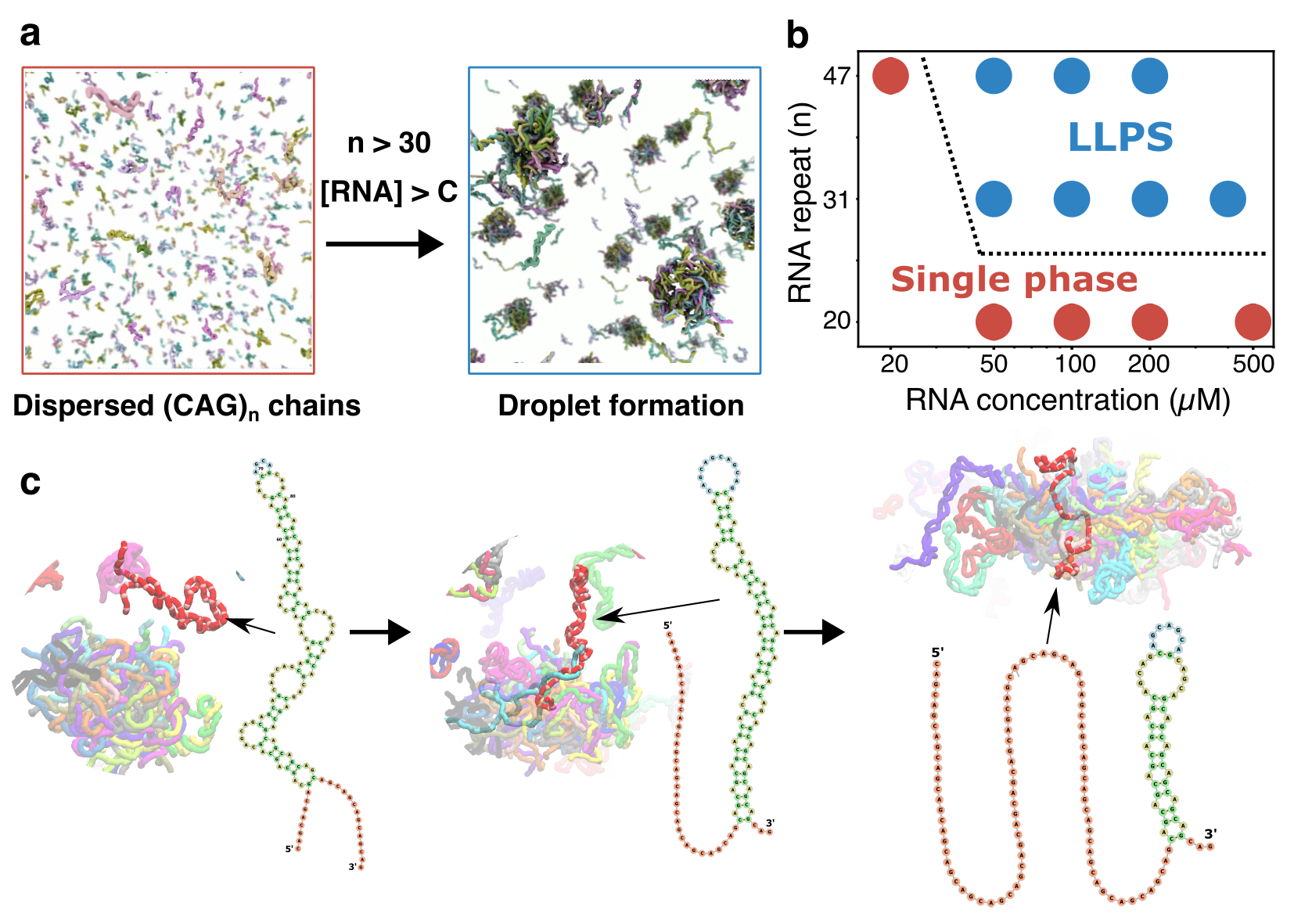}
    \caption{ RNA phase separation  using coarse-grained simulations. (a, b) The single interaction site model reproduces droplet formation (a) that occurs only when the RNA with $\sim$30 CAG repeats and RNA concentration  exceeds the critical concentration needed to cause phase separation (LLPS). (b) \cite{Nguyen22NatChem}.
    (c) Snapshots showing incorporation event of a (CAG)$_{47}$ molecule (red) into a droplet. Dramatic conformational changes occurs as can be seen in the accompanying secondary structure diagrams. \textit{Left:} Initially, the isolated (CAG)$_{47}$ chain formed a hairpin shape due to self-complementary base pairs. \textit{Middle:} The RNA is unwound from one end to form new base pairs with other (CAG)$_{47}$ chains present on the surface of droplet. \textit{Right:} Once incorporated into the droplet, each RNA chain tends to elongate as formation of inter-chain base pairs dominate due to entropic gain.}
    \label{fig:2}
\end{figure}

\section{Three Interaction Site (TIS) Model}
In the Three Interaction Site (TIS) model, each nucleotide consists of three distinct interaction sites -- phosphate (P), sugar (S), and nucleobase (B) (Fig.~\ref{fig:1}). This allows the TIS model to capture the chemically distinct behavior of these moieties. Since its first application in 2005~\cite{Hyeon05PNAS}, the TIS model has been widely used to study RNA folding in a variety of contexts. Examples include RNA folding induced by temperature jump~\cite{Hyeon08JACS} or mechanical forces~\cite{Hori16JMB}, folding pathways shifting with changes in ion concentration on the folding pathways~\cite{Roca18PNAS}, and effects of molecular crowding\cite{Denesyuk11JACS}, and solvent viscosity~\cite{Hori18JPCB}. The three-bead representation has also been widely adopted in other RNA models~\cite{Li21FrontiersMolBiosci, Tan22PLoSComputBiol} and to investigate various aspects of DNA biophysics \cite{Knotts07JCP,Chakraborty18JCTC}. 

Over the years dramatic improvements have been within the TIS model by incorporating the nearest neighbor thermodynamics into the TIS representation~\cite{Denesyuk13JPCB}. This was validated by quantitative comparison with experiments on RNA thermodynamics, such as melting of RNA hairpins and pseudoknots (PKs) (Figure 3a), which are essential building blocks of larger RNA molecules~\cite{Denesyuk13JPCB, Hori16JMB}.
Recent work has focused on developing methods to accurately describe electrostatic interactions associated with monovalent and divalent cations and phosphate groups (Figure 1b).  In most CG models, electrostatic interactions are modeled using the DH potential, which is sufficient for monovalent ions. For example, the TIS model with DH interactions accurately captured that folding pathways of a PK as a function of monovalent ion concentration, in accord with experiments \cite{Roca18PNAS}. However,  a limitation of such CG models is that the effects of divalent cations could not  be described accurately using the DH theory. This limitation was overcome in an important study that proposed a way to  include  both monovalent (Na\textsuperscript{+}, K\textsuperscript{+}, Cl\textsuperscript{-}) and divalent ions (Mg\textsuperscript{2+}, Ca\textsuperscript{2+}) in the TIS model explicitly~\cite{Denesyuk15NatChem}. 
Simulations using this model capture the ion-RNA electrostatic interactions, enabling visualization  of how binding of individual cations facilitate the self-assembly of RNA molecules~\cite{Denesyuk15NatChem,Hori21PNAS}.  

In a series of papers, the success of the TIS model with explicit ions has been demonstrated by quantitatively reproducing experiments on the folding of the \textit{Azoarcus} group I intron ribozyme \cite{Denesyuk15NatChem, Hori19BJ, Hori23NAR}.The simulations reproduced the Mg\textsuperscript{2+} concentration dependence of folding \cite{Denesyuk15NatChem}. Furthermore, specific Mg\textsuperscript{2+} binding sites agreed with the experimentally determined positions of Mg\textsuperscript{2+}~\cite{Denesyuk15NatChem, Hori19BJ} (Figure 3b). 
Interestingly, the predictive power of various other methods on Mg\textsuperscript{2+} ion distributions on the surface of RNA molecules was evaluated at the recent CASP16, reflecting the importance and difficulty of the problem. In the TIS simulations, the trajectories were dissected to predict how individual Mg\textsuperscript{2+} ions facilitate the formation of secondary and tertiary structural motifs, which cannot be tracked experimentally. The simulations also showed that subtle difference between Mg\textsuperscript{2+} and Ca\textsuperscript{2+}. Ca\textsuperscript{2+} can drive RNA folding  but not enough to reach the enzymatically active conformation ~\cite{Denesyuk15NatChem}.

Theoretical arguments and experiments have shown that RNA with a complex fold are kinetically trapped in native-like misfolded states, thus requiring considerable time to fold~\cite{Thirumalai96AccChemRes}. The TIS kinetic simulations showed that folds Azoarcus ribozyme folds by a multi-step process involving a direct path to the native state as well as trapping in misfolded metastable native-like states  upon a jump in the Mg\textsuperscript{2+} concentration~\cite{Hori23NAR}. Analyses of a series of Mg\textsuperscript{2+}-induced folding events, revealed that the fate of the RNA molecules -- whether they rapidly fold to the catalytically active state, get trapped in intermediate states, or even misfold -- occur at the very early stage coinciding with the  chain collapse arrangement that is driven by Mg\textsuperscript{2+} ion binding to specific sites (Figure 3c). Notably, the major misfolded state that was identified in the simulation was similar to the misfolded structures of \textit{Tetrahymena} ribozyme resolved by cryo-electron microscopy~\cite{Bonilla24COSB}.

In an important application of the TIS model, ions (Mg\textsuperscript{2+}, K\textsuperscript{+}, and F\textsuperscript{-}) were included explicitly to investigate the Mg\textsuperscript{2+}-dependent transition of a fluoride riboswitch folding in the presence of K\textsuperscript{+} and F\textsuperscript{-} ions \cite{Kumar23JPCB}. Kumar and Reddy investigated the folding of the aptamer domain (AD), which led them to propose a novel mechanism for the function of the fluoride riboswitch and F\textsuperscript{-} encapsulation by the cationic pocket composed of three Mg\textsuperscript{2+} ions\cite{Kumar23JPCB}.
Folding of the AD induces sequential binding of two Mg\textsuperscript{2+} ions to the phosphate groups driven in part by a transition from an outer to an inner-shell coordination through dehydration (Fig 4a). The third Mg\textsuperscript{2+} and F\textsuperscript{-} ion bind to the riboswitch in two steps. Binding of the Mg\textsuperscript{2+} results in trigonal cationic pocket. Subsequently, Mg\textsuperscript{2+} and F\textsuperscript{-} form a water-mediated ion pair (Mg-water-F). 

Let us highlight a couple of additional applications of the TIS model. Simulations of the central domain of the bacterial ribosome, performed over a broad range of  Mg\textsuperscript{2+} concentrations, reproduced a number of experimental findings quantitatively (Figure 3d) \cite{Hori21PNAS}. Strikingly, the TIS model simulations  predicted not only the multi-pathway folding of G-quadruplex folding but also characterized the structures of the intermediates, thus complementing the experimental results \cite{Ugrina24NAR}. The handful of applications show that the combination of TIS model simulations and experiments could be used to fully determine the folding of large RNAs, which currently cannot be achieved by other methods.

\begin{figure}
    \centering
    \includegraphics[width=0.75\linewidth]{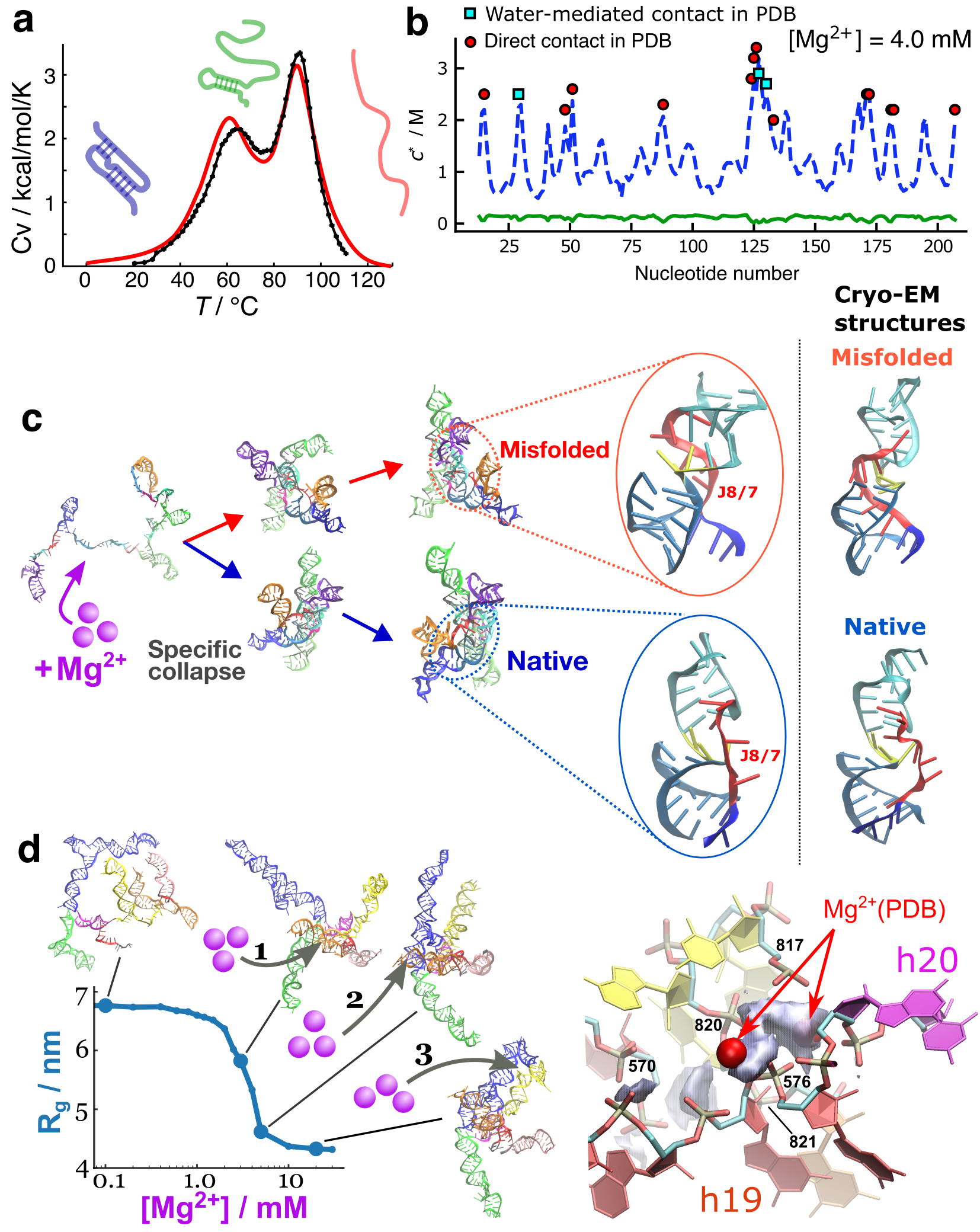}
    \caption{Ion-driven thermodynamics and kinetics of RNA folding. (a) Experimental melting profile of BWYV pseudoknot  (black)\cite{Soto07Biochem} at 500 mM NaCl and TIS simulation (red) with implicit  monovalent salt at the same concentration. (b) Nucleotide position dependent ion fingerprints in the folding of  \textit{Azoarcus} ribozyme, represented by local concentrations of Mg\textsuperscript{2+} (blue) and K\textsuperscript{+} (green).  Comparisons with the Mg\textsuperscript{2+} binding sites resolved in X-ray crystal structure are shown. (c) \textit{Left:} Kinetic partitioning mechanism in the folding of ribozyme. Differences in partial formations of secondary and tertiary structures at an early stage, driven by specific binding of Mg\textsuperscript{2+}, lead to dramatic difference in the fate -- folding to the native structure or native metastable states. A major source of misfolding is due to  topological frustration where the position of J8/7 strand (red) relative to neighboring helices are opposite to that in in the native state. \textit{Right:} The same mechanism of misfolding was recently observed in cryo-EM experiments of a similar ribozyme from \textit{Tetrahymena} \cite{Bonilla24COSB}. The images were adopted from \cite{Hori23NAR} under the CC-BY license. (d) \textit{Left:} Conformational transitions driven by Mg\textsuperscript{2+} in the central domain of bacterial ribosome. \textit{Right:} Comparison of the high density spots of Mg\textsuperscript{2+} around the central junction obtained in the simulation (grey) with the experimentally observed Mg\textsuperscript{2+} positions (red).
    }
    \label{fig:3}
\end{figure}

\section{TIS-CAT}
A technical challenge in explicitly treating both divalent and monovalent ions in simulations is that it dramatically increases the computational cost. For instance, simulating a 35-nucleotide-long pseudoknot RNA requires a system comprising 3,423 beads, 3,303 of which are for K\textsuperscript{+} and Cl\textsuperscript{-} ions. Consequently, the majority of the simulation time is spent computing interactions between monovalent ions.
To overcome this problem,  the TIS-CAT hybrid model was developed in which monovalent ions are implicitly treated using the DH potential but divalent cations are explicitly included (Figure 1b) \cite{Nguyen19PNAS}. 
The TIS-CAT model is valid when the concentration of monovalent ions (typically $>$50 mM) is higher than the divalent ion concentration, which is normally in the millimolar range in the folding experiments.
In the TIS-CAT model, we used the Reference Interaction Site Model (RISM)~\cite{Chandler72JCP}, grounded in liquid state theory in conjunction with  coarse-grained molecular simulations.
The RISM theory enables the calculations of the potential of mean force (PMF), which is the effective potential between divalent ions and the phosphate group distance that depends on the ion concentration (Figure 4a). The isotropic PMF has two distinct minima: one for inner-sphere coordination (direct binding to the phosphate) and another for outer-sphere coordination (binding through water mediation). The  dual-mode binding is crucial for accurate prediction of the thermodynamics of ion-induced RNA folding~\cite{Nguyen19PNAS}.

Simulations using the hybrid model demonstrated that divalent ions, particularly Mg\textsuperscript{2+}, have a profound effect on RNA folding through both inner- and outer-sphere coordination with phosphate groups. In smaller RNAs, such as the Beet Western Yellow Virus (BWYV) pseudoknot,  Mg\textsuperscript{2+} predominantly binds via outer-sphere interactions, where the ion remains hydrated and interacts indirectly with RNA through water molecules. In contrast, in the 58-nucleotide ribosomal RNA fragment (58-nt rRNA, Figure 4b), there is a significant shift toward inner-sphere binding, where Mg\textsuperscript{2+} directly interacts with phosphate groups after partially losing its hydration shell\cite{Nguyen20JPCB}. This change is due to the increased electrostatic potential in the more densely packed phosphate regions of larger RNA molecules, which necessitates stronger binding to stabilize the folded structure.

The simulations also highlighted the role of bridging interactions on RNA stability \cite{Nguyen20JPCB}. These occur when a single divalent ion binds simultaneously to two or more phosphate groups, acting as a bridge that stabilizes the RNA's tertiary structure. Bridging interactions are particularly important in larger RNAs where multiple phosphate groups are in proximity. 
The simulations showed that lowering the concentration of divalent ions reduces the likelihood of bridging interactions, leading to RNA destabilization. Individual nucleotides were identified as either forming or not forming interactions with Mg\textsuperscript{2+}, even in early folding or intermediate structures, depending on the bulk Mg\textsuperscript{2+} concentration.
For example, in the 58-nt rRNA, specific nucleotides were identified as key binding sites for Mg\textsuperscript{2+}, with significant accumulation of ions around them~\cite{Nguyen19PNAS}. These nucleotides are located at structurally important positions, and the Mg\textsuperscript{2+}  ions in these regions are crucial for maintaining the integrity of the folded RNA. This finding is particularly important for understanding the stability of larger RNA molecules, such as ribozymes and ribosomal RNAs, where multiple regions of the RNA come into proximity during folding.

The charge density of divalent cations plays a crucial role in determining their coordination mode with RNA. The high charge density of Mg\textsuperscript{2+}  favors outer-sphere binding in most small and intermediate-sized RNAs because it maintains a stable hydration shell~\cite{Nguyen20JPCB}. However, in regions of highly negative electrostatic potential, such as in larger RNA molecules or deep pockets of compact structures, Mg\textsuperscript{2+} can partially dehydrate and engage in inner-sphere coordination, as in the fluoride riboswitch. In contrast,  Ca\textsuperscript{2+} , which has a lower charge density and thus weaker interaction with its hydration layer, dehydrates more readily and binds directly to phosphate groups in the inner-sphere coordination, even in smaller RNAs. This difference in binding behavior is important for understanding how different divalent ions influence RNA folding and stability. The findings suggest that Mg\textsuperscript{2+} binds RNA in multiple ways depending on the RNA size and structure, while Ca\textsuperscript{2+} is more rigid in its preference for direct phosphate binding.

\begin{figure}
    \centering
    \includegraphics[width=1\linewidth]{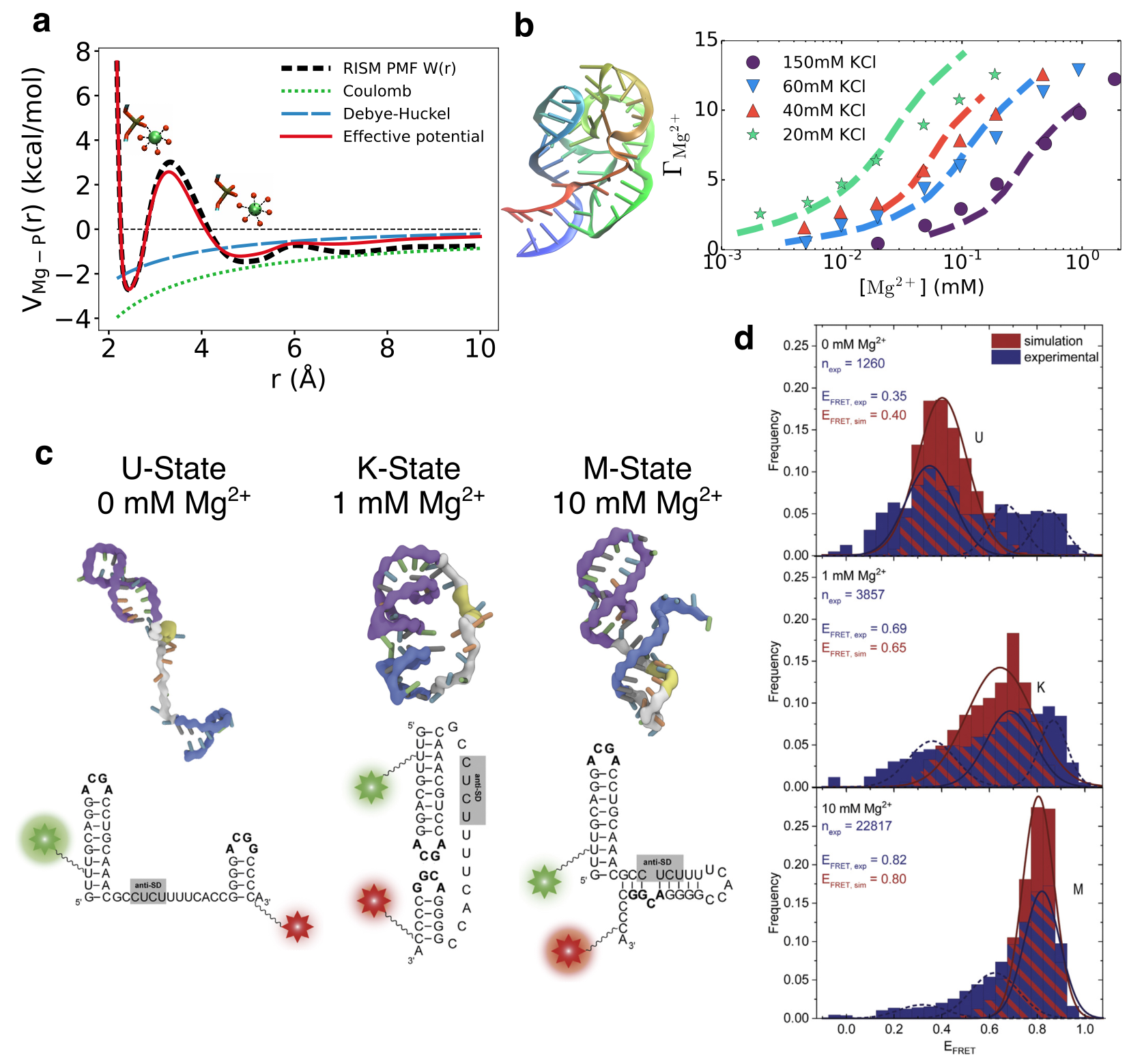}
    \caption{(a) The effective potential between phosphate and Mg\textsuperscript{2+} used in the TIS-CAT model (red) is a combination of the PMF derived from the RISM theory for the short range (grey dashed line) interaction and the Debye-H{\"u}ckel theory for the long range (cyan) potential. There are two minima corresponding to the direct (dehydrated) and water-mediated binding of Mg\textsuperscript{2+} to the phosphate. (b) Preferential interaction coefficient, $\Gamma$, depending on the bulk Mg\textsuperscript{2+} concentrations as a function of monovalent salt concentration. $\Gamma$ values calculated from TIS-CAT simulations (symbols)\cite{Nguyen19PNAS} are in excellent agreement with the experimental data (dashed curves)\cite{Grilley07Biochem}. (c) Three distinct conformations of guanidine-II riboswitch from TIS-CAT simulations at different Mg\textsuperscript{2+} concentrations (upper) and secondary structures indicating the positions of experimental FRET dies (lower) \cite{Fuks22Frontiers}. (d) Comparison of smFRET efficiency distribution between the simulations and experiments.\cite{Fuks22Frontiers} The images in (c and d) were taken from \cite{Fuks22Frontiers} under the CC-BY license \textcopyright~2022 Fuks, Falkner, Schwierz and Hengesbach.}
    \label{fig:4}
\end{figure}

Recently, single molecule FRET experiments and simulations using the TIS-CAT model were combined to detect three distinct states that are populated at different Mg\textsuperscript{2+} concentrations in guanidine-II riboswitch that controls translation 
(Figure 4c) \cite{Fuks22Frontiers}. When guanidine is bound (unbound) translation is turned on (suppressed)~\cite{Sherlock17Biochem}.  Remarkably, the calculated  distributions of FRET efficiency as a function of Mg\textsuperscript{2+} concentration using the TIS-CAT model without any adjustable parameter  were in excellent agreement with experiments (Figure 4d)\cite{Fuks22Frontiers}. The TIS-CAT simulations were used to determine the structures of the  states that are populated at three values of Mg\textsuperscript{2+} concentration. 
Note that unlike the guanidine II riboswitch, the SAM riboswitch does not suppress translation until it binds to a metabolite. Nevertheless, the CG models accurately predict the outcomes of experiments, albeit using different resolution for the RNA.

The oxRNA model, which is also based on TIS representation of RNA, has been used to calculate the melting profiles of hairpins and the MMTV pseudoknot \cite{Sulc14JCP, Matek15JCP}. The initial version~\cite{Sulc14JCP} of oxRNA did not account for electrostatic interactions. As a result, ion dependence of RNA folding was not investigated. Nevertheless, the predictions of the two melting temperatures in the MMTV pseudoknot at 1M salt were reasonably accurate.  Subsequently~\cite{Matek15JCP}, electrostatic interactions were incorporated using the DH potential. Application to supercoiling of double stranded RNA produced good agreement with experiments.

\section{Concluding Remarks}

Since the introduction of coarse-grained models to study RNA folding mechanisms twenty years ago~\cite{Hyeon05PNAS}, a great deal of progress has been made, spurred largely by advances in the experimental frontiers (such as single molecule studies, cryo-EM, time-resolved SAXS and ion-counting experiments). Although these developments, some of which are sketched here, are encouraging, the frontiers of RNA biology are rapidly moving to problems that invariably require coming to grips with complex systems involving multiple proteins, DNA, and RNA. To address such challenges, there is a clear need for more sophisticated coarse-grained models capable of accurately describing the formation of multi-component complexes, such as RNA-protein assemblies, while remaining computationally applicable to large-scale systems. One example is ribosome assembly, which has been extensively studied experimentally~\cite{Rodgers21TIBS,Shajani11ARBiochem,duss19Cell}. A normal mode analysis using elastic network representation provided insights into the dynamics of ribosomes in the ground state~\cite{Tama03NAS}, and a single-bead coarse-grained simulation captured certain motions beyond the ground state \cite{Trylska05BJ}. Nevertheless, it is fair to say that computational methods alone are currently not capable of tackling the dynamics of ribosome assembly or spliceosomes.

\bigskip

\noindent \textbf{Acknowledgments:}
This work was supported by a grant from the National Science Foundation (CHE 2320256) and the Welch Foundation through the Collie-Welch Chair (F-0019). NH was supported by JST PRESTO (JPMJPR22EA). We thank Balaka Mondal for proofreading the manuscript and the two anonymous referees who provided valuable feedback to help us improve it.

\bibliographystyle{ieeetr}
\bibliography{rna_review}

\end{document}